\documentclass[conference,10pt]{IEEEtran}
\usepackage{cite}

\usepackage[pdftex]{graphicx}

\usepackage{array}

\usepackage{url}
 
\usepackage{xspace}
\newcommand{\omnet}{{OMNeT\texttt{++}}\xspace}
\newcommand{\omnetinet}{{OMNeT\texttt{++$\:/\:$}INET}\xspace}
\usepackage[xspace]{ellipsis}		
\usepackage{microtype}				

\usepackage[colorlinks=true, urlcolor=blue, citecolor=red, linkcolor=blue, pdftex]{hyperref}

\begin{document}
\title{A New IEEE 802.15.4 Simulation Model \\ for OMNeT++ / INET}

\author{
\IEEEauthorblockN{Michael Kirsche and Matti Schnurbusch}
\IEEEauthorblockA{Computer Networks and Communication Systems Group\\
Brandenburg University of Technology (BTU) Cottbus - Senftenberg, Germany\\
eMail:~michael.kirsche@tu-cottbus.de}
}%

\IEEEspecialpapernotice{\vspace{-1.5\baselineskip}}

\maketitle

\begin{abstract}
This paper introduces a new IEEE 802.15.4 simulation model\footnote{Source code available online @ \url{https://github.com/michaelkirsche/IEEE802154INET-Standalone}} for \omnetinet.
802.15.4 is an important underlying standard for wireless sensor networks and Internet of Things scenarios.
The presented implementation is designed to be compatible with \omnet \emph{4.x} and INET \emph{2.x} and laid-out to be expandable for newer revisions of the 802.15.4 standard.
\end{abstract}


%
\section{Introduction}
\label{sec:introduction}
IEEE 802.15.4~\cite{IEEE802154} specifies a \emph{Physical} (PHY) and a \emph{Medium Access Control} (MAC) layer for use in \emph{Low Rate Wireless Personal Area Networks} (LR-WPANs).
802.15.4 itself is used as a stand-alone communication solution for resource-constrained wireless sensor nodes or as a basis for additional standards like WirelessHART or ZigBee.
It facilitates the so-called \emph{wireless embedded Internet} and \emph{Internet of Things} (IoT) scenarios in combination with the IPv6 adaptation protocol 6LoWPAN~\cite{RFC6282}.

\begin{figure}[htb]
	\centering
	\includegraphics[width=1.0\columnwidth]{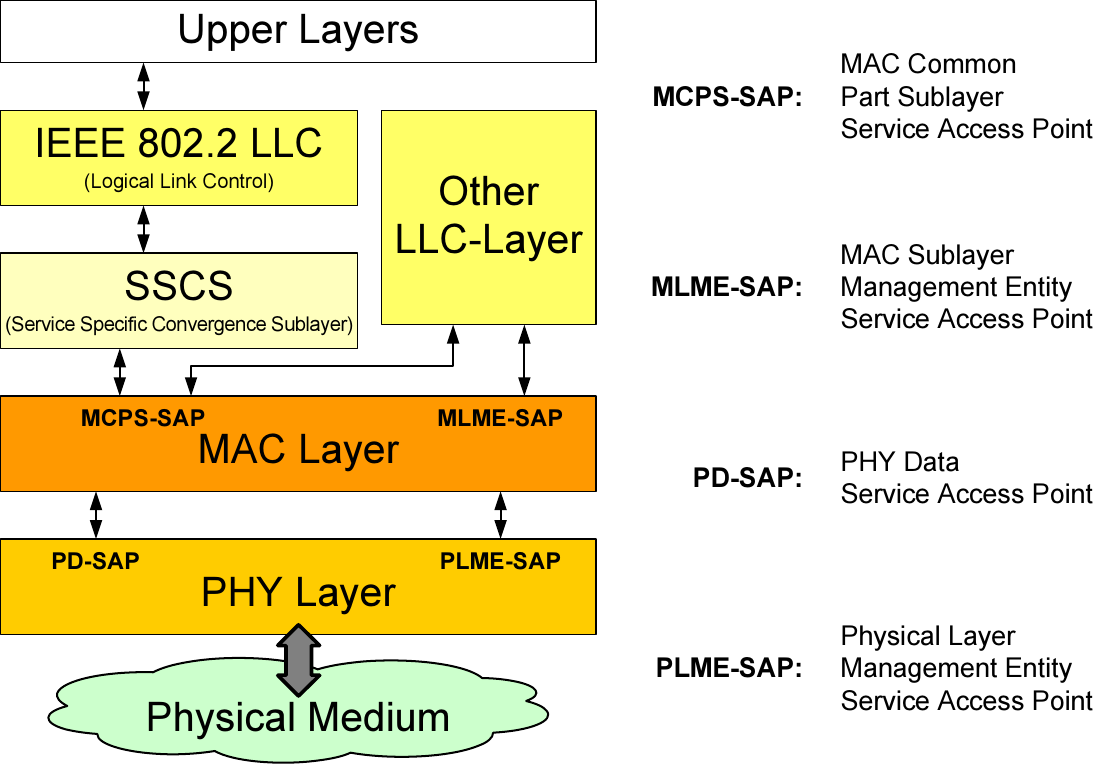}
	\caption{IEEE 802.15.4 Protocol Stack and Interfaces}
	\label{fig:IEEE802154_Stack}
\end{figure}

Layers and interfaces of 802.15.4 are depicted in Figure~\ref{fig:IEEE802154_Stack}. 
The PHY is responsible for the transceiver management (incl. frequency selection), receiving and transmitting data, as well as \emph{Energy Detection} (ED) and \emph{Clear Channel Assessment} (CCA) operations, both required for the \emph{Carrier Sense Multiple Access Collision Avoidance} (CSMA-CA) functionality of the MAC layer.
The MAC provides LR-WPAN management operations, CSMA-CA, network beacon operations, security functions, a Guaranteed Time Slot (GTS) mechanism with a superframe structure, as well as the provision of a reliable link between MAC instances.
Both layers define header and frame structures and a number of command messages.
In addition, a \emph{Service Specific Convergence Sublayer} (SSCS) is specified to adapt IEEE 802.2 LLC frames for use with the 802.15.4 MAC.
Additional descriptions are provided by \cite{IEEE802154}.
\par
802.15.4 simulation models are already included in the \omnet extension frameworks INETMANET\footnote{\url{http://github.com/aarizaq/inetmanet-2.0}}, MiXiM\footnote{\url{http://mixim.sourceforge.net/}} and Castalia\footnote{\url{http://castalia.forge.nicta.com.au}}.
Our reasons to create a new implementation for INET \emph{2.x} are manifold.
MiXiM's 802.15.4 implementation focuses on the PHY layer, with an implementation of the 802.15.4A Impulse Radio Ultra Wide-Band (IR-UWB) standard in addition to the narrow-band version. 
The MAC model only covers the CSMA-CA functionality of 802.15.4; other functions are not included.
Castalia's implementation includes an extensive PHY model and many MAC functions.
Castalia's main drawbacks are the lack of higher layer protocols for Internet applications (i.e., IPv6, TCP and UDP for use in IoT scenarios) as well as support for beacon management and newer \omnet releases.
INETMANET includes a port of an older IEEE 802.15.4 model~\cite{chen2007simulation}, originally developed for INET-2006/\omnet 3.4.
INETMANET's implementation is missing support for newer 802.15.4 revisions as well as several MAC layer service primitives, channel scan operations, and various beacon and WPAN management functions.
As the wireless embedded Internet requires protocols from the TCP/IP family, we favor using the INET framework and implementing a new 802.15.4 simulation model from scratch, to enable feature completeness as well as further combinations with other IoT standards like 6LoWPAN, CoAP and RPL.

%
\section{Implementation Details}
\label{sec:implementation-details}
\begin{figure}[htb]
	\centering
	\includegraphics[width=1.0\columnwidth]{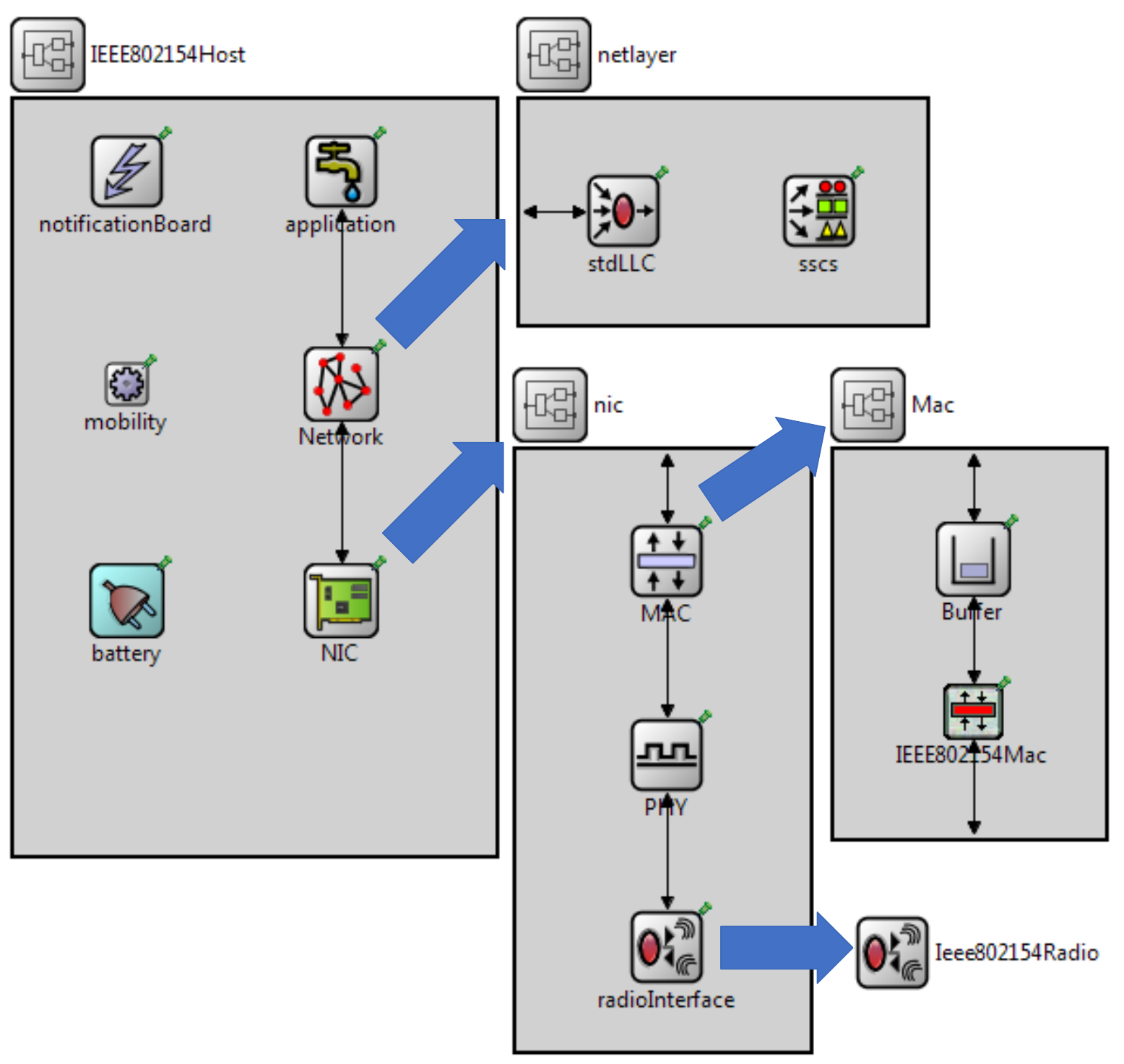}
	\caption{IEEE 802.15.4 Host with Sub-Modules}
	\label{fig:IEEE802154_Host}
\end{figure}

The composition of implemented (compound) models is shown in Figure~\ref{fig:IEEE802154_Host}.
We modeled each layer, their connecting interfaces as well as the service primitives according to the standard specifications (refer to Figure~\ref{fig:IEEE802154_Stack} and \cite{IEEE802154}) and general modeling guidelines for IEEE 802.15.4 \cite[Sec. 12.3]{wehrle2010modeling}.
\par
The 802.15.4 model itself consists of the following parts:

\begin{itemize}
	\item	\texttt{IEEE802154Radio}: This base class is an extension of INET's abstract \emph{Radio Class}. It includes the reception and the radio model with adjustments for 802.15.4 (e.g., \emph{Energy Detection} (ED), \emph{Clear Channel Assessment} (CCA) functions, link layer service primitives).
	\item	\texttt{IEEE802154PHY}: This class represents the PHY layer. Message exchange between the MAC layer and the radio interface is handled here and service primitives for CCA and ED functions are generated in this class.
	\item	\texttt{IEEE802154MAC}: The MAC layer completes the \emph{Network Interface Card} (NIC) of a IEEE 802.15.4 host. This layer manages MAC functions such as CSMA-CA, GTS, beacon and WPAN management, amongst others.
	\item	\texttt{SSCS}/\texttt{stdLLC}: Both modules connect applications to 802.15.4's MAC layer. SSCS is used for IEEE 802.2-compatible \emph{Logical Link Control} (LLC) instances (i.e., applications), whereas the \texttt{standardLLC} provides means for conversions of messages from non-802.2-compatible applications like INET's \texttt{IPvXTrafGen}.
\end{itemize}

We combined above described layers and modules into a so-called \texttt{IEEE802154Host}, similar to INET's \texttt{StandardHost}.

The following classes / headers are not included in Fig.~\ref{fig:IEEE802154_Host}, yet they are crucial for the simulation model's functionality:
\begin{itemize}
	\item	\texttt{IEEE802154Enum}: Header that describes 802.15.4's command, frame and timer types, amongst other enumerations from the standard specification.
	\item	\texttt{IEEE802154Fields}: Header that defines \emph{structs} for complex 802.15.4 data structures like the PAN descriptor or the superframe specification.
	\item	\texttt{MACAddressExt}: Extends INET's \texttt{MACAddress} class with support for EUI-64 (64-bit \emph{Extended Unique Identifier}) MAC addresses and 16-bit (IEEE 802.15.4) short addresses.
	\item	\texttt{MACFrameControlField}: Class that generates the MAC frame control field (refer to \cite[Sec. 7.2.1.1]{IEEE802154}).
	\item	\texttt{MACPIB} / \texttt{PHYPIB}: Classes that represent MAC and PHY PAN Information Bases (PIB) -- databases that hold attributes required for the layer management.
\end{itemize}

\subsection*{Modus Operandi}
The modules are initialized according to INET's multi-stage initialization procedure.
Parameters are set to their default or user-specified values (via \texttt{omnetpp.ini}).
Necessary timers for the MAC and PHY layer (i.e., the radio module) are started and the 802.15.4 node goes into receive state if nothing else is specified and no traffic generator is started.
\par
Packet reception and transmission is modeled with the use of according service primitives as specified in the standard.
Packets and command frames are transferred over their corresponding interfaces.
Different use cases (e.g., direct data transfer, indirect data transfer, GTS handling) are modeled and available as example configurations in the \texttt{omnetpp.ini}.

%
\section{Ongoing and Future Work}
\label{sec:ongoing-work}
We are currently in the process of adapting the 802.15.4 model to the latest INET version and update its specifics to recent revisions of the 802.15.4 standard.
At this stage, the model is used and developed as a standalone project. 
We plan to integrate it completely into INET, thus simplifying the use of 802.15.4 in INET-based simulations.
For this, we plan to add \texttt{ILivecycle} support and integrate the 802.15.4 modules in INET's layer structure and its \texttt{StandardHost}.
Providing an integration in \omnetinet's project feature dialog is also outstanding work.
\par
Certain modulation schemes (e.g., ASK and QPSK) are not used in the 802.15.4 model nor provided in INET yet. 
We need to tackle this issue along with the parametrization, inclusion of newer PHYs (from current 802.15.4 revisions) and the practical validation of the 802.15.4 model against real-life sensor nodes.
Choosing PHY and energy consumption parameters is another open question, as 802.15.4 hardware platforms from different vendors provide different characteristic numbers.
\par
Last but not least, we plan to combine and integrate the 802.15.4 model with our 6LoWPAN simulation model~\cite{paper+kirsche-13:6lowpan-omnet} for INET to expand the use of \omnetinet in the area of sensor networks and IoT simulation.

\bibliographystyle{IEEEtran}

%
\end{document}